\documentclass[prd,twocolumn,preprintnumbers]{revtex4}
\usepackage{amsmath}
\usepackage{graphicx}

\newcommand{\ra}{\rangle}
\newcommand{\la}{\langle}
\newcommand{\Zeff}{\ensuremath{Z_{\mathrm{eff}}}}

\begin{document}
\preprint{\fbox{LA-UR-09-04487}}
\title{Reaction-in-Flight Neutrons as a Signature for Shell Mixing in NIF capsules}
\date{\today}

\author{A.C. Hayes}
\author{P.A. Bradley}
\author{G.P. Grim}
\author{Gerard Jungman}
\author{J.B. Wilhelmy}
\affiliation{Theoretical Division, Los Alamos National Laboratory, Los Alamos, NM  87545}

\begin{abstract}
We present analytic calculations and results from computational simulations showing that
reaction-in-flight (RIF) neutrons act as a robust indicator for mixing of the ablator shell
material into the  fuel in DT capsules designed for the National Ignition Facility.
The sensitivity of RIF neutrons to hydrodynamical mixing arises 
through the dependence of RIF production on charged-particle stopping lengths in the mixture of
DT fuel and ablator material. Since the stopping power in the plasma is a sensitive function
of the electron temperature and density, it is also sensitive to mix. RIF production scales
approximately inversely with the degree of mixing taking place, and the ratio of RIF to
downscattered neutrons provides a measure of the mix fraction and/or the mixing length.
For sufficiently high-yield capsules, where spatially resolved RIF images may be possible,
neutron imaging could be used to map RIF images into detailed  mix images.
\end{abstract}

\maketitle

At the National Ignition Facility, reaction-in-flight (RIF) neutrons
are expected to make up to about 1.5\% of the total neutron production in high-yield
capsule implosions. In this paper, we show how RIF production is sensitive to mixing
of the capsule shell material into the DT fuel and how this sensitivity can be used
as a probe of such mixing.
RIF neutron production is affected by mix through a direct dependence 
on charged-particle stopping.
Because mixing causing changes in electron temperature and density, it directly
affects the charged-particle range in the plasma.
Energetic ions  undergo fewer reactions-in-flight if they are stopped more
quickly, so the number of RIF neutrons decreases with a decrease in the charged-particle range.
As discussed below, our radiation-hydrodynamic-burn
simulations confirm the expectation that the magnitude of the RIF neutron
spectrum (relative to the total neutron production) scales directly with mix. This is in contrast to the other components of the
neutron spectrum, which, to first order, simply scale up and down with the yield.

RIF neutrons are produced by a two-step process. In the first step,
a primary 14.1 MeV DT neutron knocks a triton or deuteron up to a spectrum of energies from zero
to more than 10 MeV for tritons. In the second step, the energetic knocked-on ion undergoes
a DT reaction with a thermal ion, producing a neutron above the primary 14.1 MeV peak.
The continuous spectrum of neutrons produced in this way constitute the RIF neutron spectrum. 
The spectrum of knock-on particles after transport away from the point of production
has a similar shape, but is distorted by energy loss in the DT fuel \cite{radha,petrasso}.

\begin{figure}
\includegraphics[width=2.0in]{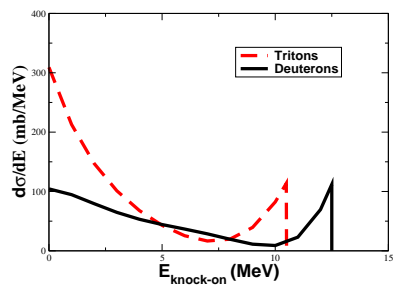}
\caption{\small Differential cross section for knock-on production by elastic scattering of 14.1 MeV neutrons
on deuterons and tritons, as a function of the energy of the scattered ion. These spectra represent the
energy distribution for knock-on ions at the point of production.
\label{fig:kospec}}
\end{figure}

A second mechanism for production of RIF neutrons begins with energetic $\alpha$ particles
born with 3.5 MeV of kinetic energy in the reaction $d + t\rightarrow n + \alpha$. These
$\alpha$ particles collide with ions, producing another distinct component of
high-energy knock-ons. RIF neutrons produced by reactions of these high-energy particles
contribute to the RIF spectrum up to about 17.6 MeV. Their production involves
two powers of the charge-particle stopping power, one for the alpha particle and one for the
knocked-on triton or deuteron.
For this reason, $\alpha$-induced RIF production shows a different sensitivity to mix than normal RIFs. 
Because the charged particle stopping is short in the DT fuel of a NIF capsule, the ratio of alpha-induced
to neutron-induced knock-on ions is small. For the rest of this paper, we concentrate on neutron-induced
RIFs only.

The mix dependence of RIF production comes into play in
the transport and energy loss of the energetic deuteron or triton, after
it is knocked on. This charged-particle
transport is sensitive to the plasma conditions\cite{radha, petrasso}; in the
relevant temperature regime (keV), charged-particle stopping power
is dominated by scattering of ions with electrons in the plasma and behaves like\cite{Li, singleton}
\[
  -\frac{dE}{dx} \simeq c \frac{n_e}{\theta^{3/2}}\log(\Lambda) E^{1/2},
\]
where \(\theta\), \(n_e\) are the electron temperature and
number density and $E$ is the kinetic energy of the ion; $\log(\Lambda)$ is the  Coulomb logarithm
that describes the interplay between the long- and short-range Coulomb interactions,
and \(c\) is a constant very nearly independent of plasma conditions. In principle, the energy distribution
of charged-particles at a point distant from their point of production can be calculated directly
in terms of the distribution at the point of production,
using this stopping power.
The rate of RIF reactions at any point is the product of the deuteron/triton density at that point,
the cross-section for $d + t \rightarrow \alpha + n$, and the flux of transported knock-on ions
at that point. The total RIF production over the whole capsule is the integral of this local rate
and clearly depends directly on the charged-particle stopping length and the processing of
the knock-on production spectrum by charged-particle transport.

Formally, the local RIF production rate is defined by an integral over charged-particle
paths, from the knock-on production point to the RIF reaction point. Let $y$ be the point of
production for a knock-on ion, and let $x$ be the RIF reaction point. The path
is the segment connecting $y$ to $x$. Suppressing the energy
dependence, we can write
\begin{equation}
  \frac{d \Gamma_{\mathrm{RIF}(x)}}{dV} =
  \frac{1}{2} n_{dt}(x) \sigma_{ko} \sigma_{dt} \int dy\, \phi_n(y)\, n_{dt}(y)\, \psi_{ko}(y, x),
\label{rif-eq}
\end{equation}
where \(\phi_n\) is the primary neutron flux and \(\psi_{ko}\) is the resulting
knock-on fluence (per unit initial particle), and \(\sigma_{ko}\) and \(\sigma_{dt}\) are
cross sections determining the knock-on and $d+t$ reaction rates. 
Evaluating eq. (\ref{rif-eq}) requires, in general, 
detailed simulations that include  treatment 
of the charged-particle transport.

Before turning to our simulations, we note that considerable insight into the physics
of RIF production can be obtained by expressing the charged-particle transport
in terms of average paths for neutrons and for knock-on particles.
Without loss of generality, the total number of charged-particle knock-ons
produced in the capsule can be written
\[
  N_{ko} = N_{\mathrm{primary}} \la n_{dt}\;\ell_{n}\; \sigma_{ko} \ra_{n},
\]
where $N_{\mathrm{primary}}$ is the number of 14.1 MeV neutrons from the primary $d+t$ reactions;
the notation $\la \cdots \ra_{n}$ is short-hand for the integral over neutron paths, averaged
over all paths taken by the $d+t$ neutrons produced in the capsule. Because the capsule is
thin to neutrons, the average length of neutron paths is, up to dimensionless geometric
factors, equal to the size of the system. In a system thick to neutrons, the average
length would be equivalent to the neutron mean free path.
Similarly, the number of RIFs is
\[
  N_{\mathrm{RIF}} = N_{ko} \la \bar{n}_{dt}\;\ell_{ko}\;\sigma_{dt}\; \ra_{ko},
\]
where $\bar{n}_{dt}= \frac{1}{2}n_{dt}$ is the density of either D or T ions, depending
on whether the knock-on particles are T or D ions, respectively. We assume
equal populations of D and T for our capsules of interest, which leads to the
explicit factor of $\frac{1}{2}$ in this definition.
Clearly the lengths of knock-on particle paths scale directly with the charged-particle
stopping length, since the stopping length is small compared to the size of the system,
for NIF capsules.

Continuing with the qualitative discussion, we can factor the above expressions into
products of averages, each with the correct dimensions, and each scaling appropriately
with changes in the underlying plasma properties.
The number of RIF reactions taking place is then
\begin{eqnarray}
  N_{\mathrm{RIF}}
    &\approx& \frac{1}{2} N_{\mathrm{primary}}\sigma_{ko}\sigma_{dt} \;n_{dt}^2\; R \; \ell_{ko} \\
    &\approx& \frac{1}{2} N_{\mathrm{primary}}\sigma_{ko}\sigma_{dt} \;n_{dt} R \; \frac{n_{dt}}{n_e}\; \theta_e^{3/2} \; \frac{1}{c \log(\Lambda)},
\label{rifapprox}
\end{eqnarray}
where \(R\) is the size of the system.
In a limit where the plasma is thin to charged-particles, $N_{RIF}$ would scale with $(\rho r)^2$ or,
equivalently, with $n^2_{dt} R^2$. However, NIF capsule implosions of interest are not
in this regime. As indicated above, when the charged-particle stopping length is short compared
to the size of the system, $N_{RIF}$ scales with only one power of $n_{dt}$.
The second power of $n_{dt}$ only appears as a ratio to the electron density, and
charge neutrality requires that the ratio of ions to electrons \(n_i/n_e\) is
a constant dependent only on the effective charge of the ions; to first order, variations
in this ratio only occur with changes in material composition, such as those which occur due to mixing.

\begin{table}
\caption{Definition of the mix fraction $f$ in terms of the ion and/or electron densities in the gas. \label{tab:fractions}}
\begin{tabular}{|l|l|}
\hline
density of ions in the gas  & $n_{gas}=n_{dt}+n_m$      \\
density of mix ions         & $n_m=f\;n_{gas}$          \\
density of DT ions          & $n_{dt}=(1-f)\;n_{gas}$   \\
                            &                           \\
electron density            & $n_e=n_{dt} + \Zeff n_m$  \\
                            &$ n_e=(1-f +\Zeff)n_{gas}$ \\
\hline
\end{tabular}
\end{table}

Therefore, we next consider the effect of mixing on RIF production in this simple formulation.
Using the definitions of mix fraction $f$ in Table \ref{tab:fractions}, Eqn. (\ref{rifapprox}) implies that 
the change in RIF production, relative to a clean no-mix ($f=0$) situation, is simply
\begin{equation}
\frac{(N_{RIF}/N_{primary})_{MIX}}{(N_{RIF}/N_{primary})_{CLEAN}}
   = \frac{(1-f)^2}{1-f +\Zeff\; f}\; \left(\frac{\theta_e^{mix}}{\theta_e^{DT}}\right)^{3/2}
\label{eqn:transfer}
\end{equation}
$\theta_e^{mix}$ is the electron temperature in the presence of mix and $\theta_e^{DT}$ 
is the  temperature in the unmixed fuel. $\Zeff$ is the number of free electrons introduced into the
gas per mix atom.
Going beyond these assumptions, which lead to a single representative charged-particle stopping length,
requires realistic simulations, such as those  presented below.
We note that the function describing the dependence of the RIF fraction on the  mix fraction $f$ 
is sometimes referred to as the  {\it transfer function}.

\begin{figure}[ht]
\includegraphics[width=2.75in, angle=-90]{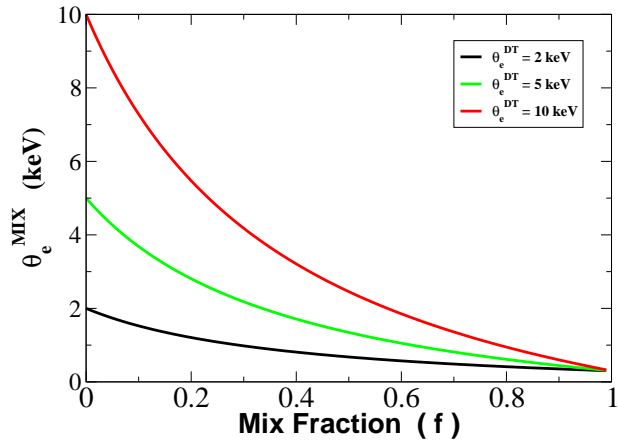}
\caption{\small (color online)
The temperature in the mix region as a function of mix fraction f.
In these calculations, $Z_{\mathrm{eff}}$ was taken to be 6. The unmixed shell material
temperature was assumed to 300 eV, which is a typical value obtained in simulations.
The different curves are for three different DT electron temperatures,
as indicated in the legend.}
\label{fig:pic1}
\end{figure}

To further fix the parameters of this qualitative discussion, we can consider models which
couple the change in electron temperature to the mix fraction, thereby fixing the
ratio in Eqn. (\ref{eqn:transfer}) explicitly in terms of the mix fraction $f$.
The simplest model of this type, which is not far removed from the true physical
situation, stipulates an adiabatic mixing process.
In the adiabatic limit, in which heat is conserved, the  mix-induced change in the electron temperature 
can be related to the electron temperatures in the original un-mixed shell and un-mixed DT fuel regions by
\begin{equation}
\theta_e^{mix} =  \frac{(1+\Zeff)f \theta_e^{shell} + 2(1-f)\theta_e^{DT}}{(1+\Zeff)f + 2(1-f)}\;\;.
\label{eqn:temp}
\end{equation}
This change in temperature as a function of mix fraction is shown in Fig. \ref{fig:pic1}.

\begin{figure}[ht]
\includegraphics[width=2.75 in, angle=-90]{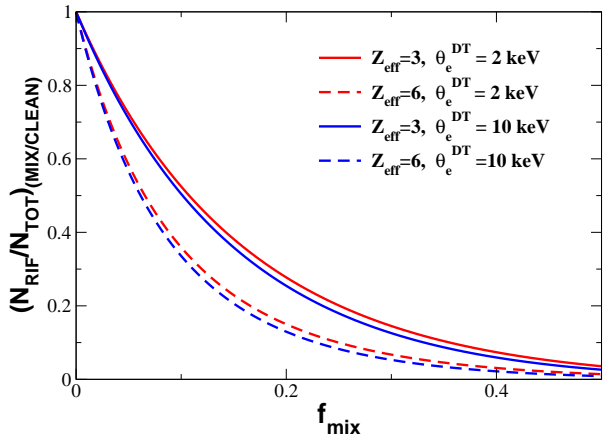}
\caption{\small(color online) RIF production, compared to a clean calculation, as a function of the mix fraction $f$
for different values of  $\Zeff$, the number of free electrons introduced into the gas per
mix  atom.  The curves shown are derived from eqs. (\ref{eqn:transfer}-\ref{eqn:temp})
which makes the over-simplified assumption that the material properties do not vary in
the RIF production region. The temperatures $\theta_e^{DT}$ are the assumed temperatures of
the electrons in the DT fuel without mix ($f_{mix}=0$).  }
\label{RIF-dumb}
\end{figure}

Combining eqs. (4) and (5), the expected shape of the RIF production rate with increasing mix fraction $f$ is as shown in
Fig. \ref{RIF-dumb}. As can be seen, the rate of suppression of the RIF production with increasing
mix fraction is controlled by $\Zeff$. Of course, in reality it is unlikely that there would be a single mix fraction or
$\theta_e^{mix}$ throughout the capsule; rather, $f\equiv f(x,y,z)$ in cylindrical coordinates. 
For high-yield shots, the RIF production is expected to be large enough to allow 
for spatially resolved RIF neutron images\cite{grim-to-be}.
Using Eqs. 4 and 5, (or, equivalently, Fig. 2),  a RIF neutron image
could be translated into a spatially resolved map of the mix fraction.
In other words, the so-called transfer function 
determines the mapping needed to convert a two-dimension neutron image of RIF/Total$(x,y)$ into an
$f(x,y)$ mix image.
Such a capability would open the possibility of 
detailed hydrodynamical mixing experiments at the NIF, assuming that yields sufficient for 
 a {\it Uses of Ignition} program are obtained.
  
Equations \ref{eqn:transfer} and \ref{eqn:temp} were derived under the assumptions that complications
from charged-particle transport and capsule dynamics could be ignored.
To examine the changes in RIF production with increasing mix fraction under more realistic conditions,
we ran a series of one-dimensional radiation-hydrodynamic-burn simulations. 
Hydrodynamic mixing was treated using 
the Scannapieco and Cheng mix model\cite{scan}, which is essentially a
fluid interpenetration model.  In this model, the degree of mixing is controlled by
a single parameter $\alpha$, which controls the
dynamic mixing length. 
Analyses \cite{wilson-mix} of Omega experiments typically require a value of $\alpha\sim 0.06$
in order to reproduce  observed yields. In the present calculations, we allowed mixing across
both the shell/ice and ice/gas interfaces. We varied $\alpha$ between 0-0.16, but always used
the same value of $\alpha$ at the two interfaces.
Charged-particle transport was calculated using an effective diffusion model.
We have also conducted tests of the basic physics of the RIF production mechanism outlined
in this paper, under more simplified physical circumstances, but using highly accurate Monte Carlo
charged-particle transport.
More accurate Monte Carlo transport methods could change the results of the full-physics
simulations quantitatively, but will not change the conclusions of the paper.

\begin{figure}[h]
\includegraphics[width=3.5in]{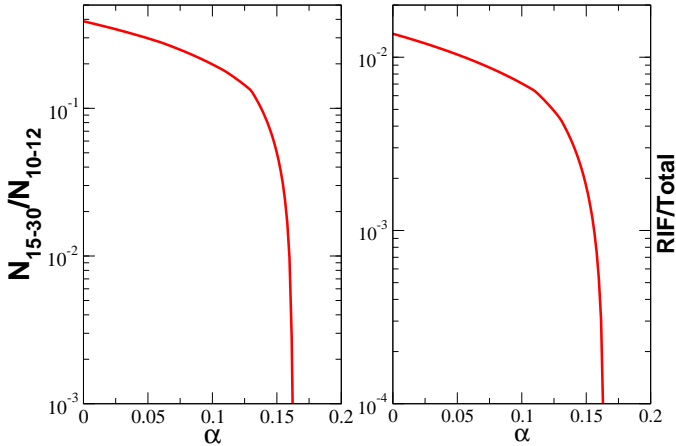}
\caption{\small The panel on the left shows the RIF to downscattered neutron ratio with increasing mixing
of the shell material into the fuel. The panel on the right is the RIF to total neuron ratio.
$\alpha$ is the parameter in the
Scannapieco and Cheng model that controls the mixing length.
Here the RIF neutrons are  all neutrons above 15 MeV,
the downscattered are neutrons in the energy
range E$_n$=10-12 MeV,  the total  are all those
in the range 0-30 MeV.  
}
\label{rif-down}
\end{figure}

Since the RIF production depends on the areal density of the fuel, the RIF dependence on mix is
most simply displayed by dividing out the areal density. For this reason we examined the ratio of the
RIF neutrons to  downscattered neutrons, since the latter are a direct measure of the fuel
$\la\rho r\ra$ \cite{wilson, azechi}. Our definition of RIFs included all
neutrons above 15 MeV, while the downscattered neutrons included all neutrons
in the energy range 10-12 MeV. 
The energy restriction on the latter range avoids
complications from neutrons produced in $t+t$ and other reactions.
Choosing different energy ranges for either the RIF or downscattered neutrons did not significantly
change the sensitivity of RIFs to mix.
The results are shown in Fig. \ref{rif-down}, where  a steady drop is seen 
in the RIFs to downscattered ratio as the mix is increased.
The decrease in the RIF ratio scales approximately linearly with the mixing length,
until the degree of mixing is sufficiently large to induce  failure.
\vspace*{1.25cm}

In summary, both analytic arguments and computational simulations 
show that RIF neutrons act as a robust signature for 
mix in NIF capsules. 
The RIF neutrons could be measured either by neutron time-of-flight \cite{Gle06} or
by using radiochemical techniques\cite{Hayes, Grim, bradley}.
Neutron imaging could also be used to measure RIF in high-yield capsules, and
such images could be used to obtain spatially resolved mix images.
Finally, we note that the shape of the RIF spectrum is determined in a relatively
straightforward way from the shape of the knock-on spectrum, 
and the subsequent energy loss of the knock-on charged-particles in the plasma. 
If $\alpha$-RIFs are included, the lower energy portion of the RIF spectrum
($E_n\sim 15-18$ MeV) has a more complicated shape, and
this can be used to extract additional information on mix and electron temperature.

\end{document}